 \documentclass[final,3p,times]{elsarticle}

\usepackage{epsfig}
\usepackage{amssymb}
\usepackage{amsthm}
\usepackage{amsfonts,amsmath,amssymb}

\def\qed{\ifmmode\hbox{\hfill\sqb}\else{\ifhmode\unskip\fi%
\nobreak\hfil
\penalty50\hskip1em\null\nobreak\hfil\sqb
\parfillskip=0pt\finalhyphendemerits=0\endgraf}\fi}

\newcommand{\R}{\mathbb{R}}
\newcommand{\N}{\mathbb{N}}
\newcommand{\E}{\mathbb{E}}

\newtheorem{thm}{Theorem}
\newtheorem{lem}[thm]{Lemma}
\newdefinition{rmk}{Remark}
\newproof{pf}{Proof}

\journal{Statistics \& Probability Letters}

\begin{document}

\begin{frontmatter}

\author[rvt]{L. Bordes\corref{cor2}}
\ead{laurent.bordes@univ-pau.fr}
\author[rvt]{C. Paroissin}
\ead{cparoiss@univ-pau.fr}
\author[rvt]{A. Salami}
\ead{ali.salami@univ-pau.fr}

\cortext[cor2]{ http://lma-umr5142.univ-pau.fr, T\'el. 05 59 40 75 38, Fax  05 59 40 75 55} 

\address[rvt]{Universit\'{e} de Pau et des Pays de l'Adour, Laboratoire de Math\'{e}matiques et de leurs Applications - UMR CNRS 5142, Avenue de l'Universit\'{e}, 64013 Pau cedex, France.}

\title{Parametric inference in a perturbed gamma degradation process}

\begin{abstract}
We consider the gamma process perturbed by a Brownian motion (independent of the gamma process) as a degradation model. Parameters estimation is studied here. We assume that $n$ independent items are observed at irregular instants. From these observations, we estimate the parameters using the moments method. Then, we study the asymptotic properties of the estimators. Furthermore we derive some particular cases of items observed at regular or non-regular instants. Finally, some numerical simulations and two real data applications are provided to illustrate our method.
\end{abstract}

\end{frontmatter}

\section{Introduction and model}
\label{para1}
Many authors model degradation by a Wiener diffusion process. Doksum and H\'oyland \cite{Dok} applied the Brownian motion with drift to a variable-stress accelerated life testing experiment.  Next Whitmore \cite{Whi1} extended the Wiener degradation process with the possibility of imperfect inspections.  Another interesting extension is the bivariate Wiener process considered by Whitmore et al.\cite{Whi2} in which the degradation process and a marker process (that can be seen  as a covariate in medical applications) are combined. Finally Wang \cite{Wang} has studied the maximum likelihood inference method for a class of Wiener processes including random effects. According to Barker \cite{Bar}, this process is no longer monotone, but can take into account minor system repairs over time. In addition, this process can be negative. Although such behaviours have difficult physical interpretation. They can be explained by above mentioned phenomena like minor repairs or measurement degradation errors. It means that for some types of degradation models, the possibility of non-negative increments is appropriate.

However in many situations the physical degradation process can be considered as monotone while the observed process is a perturbation of the degradation process and then can be no longer monotone. Physical degradation processes are usually described by monotone L\'{e}vy processes like the gamma process or the compound Poisson process. These process implies that the system state cannot be improved over time, and then this system cannot return to its original state without external maintenance actions. The gamma process was originally proposed by Abdel-Hameed \cite{Hameed} in order to describe the degradation phenomenon. This process is frequently used in the literature since it is preferable from the physics point of view (monotonic deterioration). Moreover, calculations with this process are often explicit, it properly accounts for the temporal variability of damage and allows determining optimum maintenance policies

In this paper, we propose a degradation model $D=\left( D_{t}\right)_{t\geq 0}$ which combines these two approaches as follows:
$$ \forall t \geq 0\,,\,D_{t}=Y_{t}+\tau B_{t}$$
where $(Y_{t})_{t \geq 0}$ is a gamma process such that $Y_{1}$ is gamma distributed with scale parameter $\xi > 0 $ and shape parameter $\alpha > 0$ and where $(B_{t})_{t \geq 0}$ is a Brownian motion. This model is defined for $\tau \in \R$ and the two processes are assumed to be independent. Without loss of generality, we can assume that $\tau \geq 0$ since $\tau B_{t}$ and $-\tau B_{t}$ have the same distribution for all $t \geq 0$. The motivations behind considering such a model are the following ones. First, this model embeds the two approaches mentioned above. Indeed, it is clear that when $\tau = 0$, this model turns to be a gamma process. Moreover, if $\alpha/\xi$ tends to $b>0$ and $\alpha/\xi^{2}$ tends to $0$, then this model converges weakly to a Brownian motion with positive drift $b$. Second, measurements of degradation tests reflect measurement errors. Hence, the role of Brownian motion in this model can be interpreted as measurement errors. Finally, our model can take into account minor repairs considered on system over time.

In this paper, estimation of model parameters is derived using the method of moments.  In literature the two most common methods of gamma process parameter estimation, namely, maximum likelihood and method of moments, are discussed in \cite{vanour}. Both methods for deriving the estimators of gamma process parameters were initially presented by \c Cinlar et al.\cite{Cin}. Besides, Dufresne et al.\cite{Duf} propose to use a conjugate Bayesian analysis in which the scale parameter of the gamma process is assumed to have an inverted gamma distribution as prior. A method for estimating a gamma process by means of expert judgement and a Bayesian estimation method is also discussed in \cite{vanour}.  Finally maximum-likelihood and Bayesian estimation of the parameters of the Brownian stress--strength model was studied by Ebrahimi and Ramallingam \cite{Ebra} and Basu and Lingham \cite{Basu}.

The organization of the paper is as follows. First, we present a general case where $n$ independent processes are observed at irregular instants. Both number of observations and instants are different for each degradation process. Parameters estimation and asymptotic properties (consistency and asymptotic normality) of the estimators are studied. Next, we derive some particular cases of items observed at regular or non-regular instants. Finally, numerical simulations and two real data applications are provided to illustrate our method.

\section{General case}

Let $\left( D^{(n)}\right)_{n \in \N^{\ast}}$ be a sequence of independent and identically distributed (i.i.d.) copies of the degradation model described in the previous section. The $i$-th degradation process is observed $N_{i}$ times such that $N_{i} \in \N^{*}$. For all $i \in \N^{\ast}$ and all $j \in \left\lbrace 0, \ldots, N_{i}\right\rbrace$, we will denote by $t_{ij}$ these instants (with convention that for all $i \in \N^{\ast}$, $t_{i\,0} = 0$). Let $\theta = \left(\xi, \alpha, \tau^{2}\right) \in \Theta = \R^{*}_{+}\times \R_{+}^{*} \times \R_{+}$ the parameter space of the model. Estimation of model parameters is derived using the method of moments. Asymptotic properties are then studied.

\subsection{Parameter estimation}
 
For any $i \in \N^*$, for any $1 \leq j \leq N_{i}$ and for any $k \in \N$, we denote by $m^{(k)}_{ij}$ the $k$-th moment and by by $\overline{m}^{(k)}_{ij}$ the $k$-th central moment of increments $ \Delta D_{ij} = D^{(i)}_{t_{ij}}-D^{(i)}_{t_{ij-1}}$ :
$$
m^{(k)}_{ij} = \E\left[\Delta D_{ij}^{k}\right] \quad {\mbox{and}} \quad  \overline{m}^{(k)}_{ij} = \E\left[ \left( \Delta D_{ij} - \E\left[ \Delta D_{ij}\right] \right)^{k} \right].
$$

Since the gamma process and the Brownian motion are independent, the first three moments are equal to:
\begin{eqnarray*}
m^{(1)}_{ij} & = & \frac{\alpha}{\xi} \Delta t_{ij},\\
m^{(2)}_{ij} & = &  \frac{\alpha}{\xi^{2}} \Delta t_{ij} + \left(\frac{\alpha}{\xi}\Delta t_{ij}\right)^{2}  + \tau^{2} \Delta t_{ij},\\
m^{(3)}_{ij} & = & \frac{2 \alpha}{\xi^{3}} \Delta t_{ij} + 3 \frac{\alpha^{2}}{\xi^{3}} \left(\Delta t_{ij} \right)^{2}  + \left(\frac{\alpha}{\xi}\Delta t_{ij}\right)^{3} + 3 \frac{\alpha \tau^{2}}{\xi} \left(\Delta t_{ij}\right)^{2}.
\end{eqnarray*}

These expressions can be easily computed from the moments of the gamma distribution (see \cite{Bala} for non-central moments and see \cite{Will} for a recursive formulae of the central moments) and from the ones of the normal distribution \cite{Bala}.

Let $f$ be the following differentiable map from $\Theta$ to $f(\Theta)$ defined by:
$$\forall \theta \in \Theta,\; f(\theta) =\begin{pmatrix}
   m^{(1)}\\
   \overline{m}^{(2)}\\
   \overline{m}^{(3)}
\end{pmatrix} = \begin{pmatrix} m^{(1)}_{ij}/\Delta t_{ij} \\
\overline{m}^{(2)}_{ij}/\Delta t_{ij}\\
\overline{m}^{(3)}_{ij}/\Delta t_{ij}
\end{pmatrix} = \begin{pmatrix} \alpha /\xi \\
\alpha/\xi^{2} + \tau^{2}\\
2\alpha/\xi^{3}
\end{pmatrix}.$$

The function $f$ is bijective. Then, the parameters can be expressed in terms of $m^{(1)}, \overline{m}^{(2)}$ and $\overline{m}^{(3)}$ as follows:
$$f^{-1}(m) = \begin{pmatrix}
   \xi\\
  \alpha\\
   \tau^{2}
\end{pmatrix}
=
\begin{pmatrix}
\sqrt{\frac{2 m^{(1)}}{\overline{m}^{(3)}}}\\
 m^{(1)} \sqrt{\frac{2 m^{(1)}}{\overline{m}^{(3)}}}\\
 \overline{m}^{(2)} - \frac{\sqrt{2 m^{(1)}\overline{m}^{(3)}}}{2}
\end{pmatrix}, \; \mbox{where} \; m \; = \; \begin{pmatrix}
m^{(1)}\\
\overline{m}^{(2)}\\
\overline{m}^{(3)}
\end{pmatrix}.$$

Let $\widehat{m}_{n}$ be the empirical estimator of the first central three moments:
$$\widehat{m}_{n} = \begin{pmatrix}
   \widehat{m}_{n}^{(1)}\\
   \widehat{\overline{m}}_{n}^{(2)}\\
  \widehat{\overline{m}}_{n}^{(3)}
\end{pmatrix}\; =\;\left( \sum_{i=1}^{n}N_{i}\right)^{-1}\sum_{i=1}^{n}\sum_{j=1}^{N_{i}}\begin{pmatrix} \Delta D_{ij}/\Delta t_{ij}\\
\left(  \Delta D_{ij} - \Delta t_{ij} \widehat{m}_{n}^{(1)} \right)^{2}/\Delta t_{ij}\\
\left( \Delta D_{ij} - \Delta t_{ij} \widehat{m}_{n}^{(1)}  \right)^{3} /\Delta t_{ij}\end{pmatrix}.$$

The estimator $\widehat{\theta}_{n} = f^{-1}\left(\widehat{m}_{n}\right)$ of $\theta = \left( \xi,\alpha, \tau^{2}\right)$ is therefore defined by:
\begin{eqnarray*}
\widehat{\xi}_{n} & = & \sqrt{\frac{2 \widehat{m}_{n}^{(1)}}{\widehat{\overline{m}}_{n}^{(3)}}}\;,\; \widehat{\alpha}_{n}\; = \; \widehat{m}_{n}^{(1)} \sqrt{\frac{2 \widehat{m}_{n}^{(1)}}{\widehat{\overline{m}}_{n}^{(3)}}}\;\; \mbox{and} \;\widehat{\tau}_{n}^{2}\; = \; \widehat{\overline{m}}_{n}^{(2)} - \frac{\sqrt{2 \widehat{m}_{n}^{(1)} \widehat{\overline{m}}_{n}^{(3)}}}{2}\;.
\end{eqnarray*}

\subsection{Asymptotic properties }

We first recall the following theorem (for more details see Theorem $6.7$ in \cite{Pet}).
\begin{thm}\label{thmpetrov}
Let $\left( {\mbox{a}}_{n}\right)_{n \geq 1}$ be a sequence of positive numbers. Let $\left(X_{n}\right)_{n \geq 1}$ be a sequence of independent random variables. We set $S_{n} = \sum\limits_{k=1}^{n}X_{k}$. If ${\mbox{a}}_{n} \xrightarrow[n\rightarrow\infty]{} \infty$ and $\sum\limits_{n=1}^{\infty} \left(\mbox{Var} \left[ X_{n}\right] /{\mbox{a}}_{n}^{2} \right) < \infty$, then $\left(S_{n} - \E\left[S_{n}\right]\right)/{\mbox{a}}_{n} \xrightarrow[n\rightarrow\infty]{a.s.} 0.$
\end{thm}

Then we establish the following lemma.
\begin{lem}\label{bordes}
We have that $\sum\limits_{n \geq 1}N_{n}\left( \sum\limits_{i=1}^{n}N_{i}\right)^{-2} \; < \;\infty.$
\end{lem}
 
\begin{pf}
We set $A_{n} = N_{1} + \ldots + N_{n}$. One can note that $N_{n} = A_{n} - A_{n-1}$. Then it follows that
\begin{eqnarray*}
&&\sum\limits_{k=1}^{n}\frac{N_{k}}{\left(N_{1} + \ldots + N_{k}\right)^{2}}\\
 & = & \sum\limits_{k=1}^{n} \frac{A_{k} - A_{k-1}}{A_{k}^{2}} \; = \; \frac{N_{1}}{N_{1}^{2}} +\sum\limits_{k=2}^{n} \frac{A_{k} - A_{k-1}}{A_{k}^{2}} \; \leq \; \frac{1}{N_{1}} + \sum\limits_{k=2}^{n} \frac{A_{k} - A_{k-1}}{A_{k}A_{k-1}}\; \leq \; \frac{1}{N_{1}} + \sum\limits_{k=2}^{n} \frac{1}{A_{k-1}} - \sum\limits_{k=2}^{n} \frac{1}{A_{k}} \leq  \frac{1}{N_{1}} + \sum\limits_{k=1}^{n-1} \frac{1}{A_{k}}\\
& - & \sum\limits_{k=2}^{n} \frac{1}{A_{k}}\; \leq \; \frac{1}{N_{1}} + \frac{1}{A_{1}} - \frac{1}{A_{n}} \; \leq \; \frac{2}{N_{1}} \; \leq \; 2.
\end{eqnarray*}
\hfill$\Box$
\end{pf}

In the sequel we will prove the consistency of $\widehat{\theta}_{n}$.
\begin{thm}\label{thmconsi}
Under the following assumptions :
\begin{itemize}
\item[$(\mbox{H}_{1})$] $\sum\limits_{n \geq 1}\sum\limits_{j=1}^{N_{n}}\left(\Delta t_{nj}\right)^{-1}\left( \sum\limits_{i=1}^{n}N_{i}\right)^{-2} \; < \;\infty$,

\item[$(\mbox{H}_{2})$] $ \exists \; d_{u}$, $\forall i \in \N^{\ast}$, $\forall j \in \left\lbrace 1, \ldots, N_{i}\right\rbrace $, $\Delta t_{ij} \leq d_{u}$ \;,
\end{itemize}
$\widehat{\theta}_{n}$ converges almost surely to $\theta$ as $n$ tends to infinity.
\end{thm}

\begin{pf}
One has to prove that $\widehat{m}_{n}$ tends to $m$ a.s. as $n$ tends to infinity. Indeed, since $f^{-1}$ is continuous on $f\left(\Theta\right)$, we obtain, by applying the continuous mapping theorem \cite{vander}, that $\widehat{\theta}_{n} \xrightarrow[n\rightarrow\infty]{a.s.} \theta.$ Hence let us prove the almost sure convergence of $\widehat{m}_{n}$. 

\paragraph{{\bf Almost sure convergence of $\widehat{m}_{n}^{(1)}$ to $m^{(1)}$}}
By applying Theorem \ref{thmpetrov}, it holds that:
$$\widehat{m}_{n}^{(1)} - m^{(1)}\; = \; \left( \sum\limits_{i=1}^{n} N_{i}\right)^{-1} \sum\limits_{i=1}^{n} \sum_{j=1}^{N_{i}}  \left( \frac{\Delta D_{ij}}{\Delta t_{ij}} - m^{(1)}\right)\; = \; \frac{1}{n} \sum\limits_{i=1}^{n} \underbrace{n\left(\sum\limits_{i=1}^{n} N_{i}\right)^{-1} \left[ \sum\limits_{j=1}^{N_{i}} \left( \frac{\Delta D_{ij}}{\Delta t_{ij}} - m^{(1)}\right) \right]}_{X_{i}} \xrightarrow[n\rightarrow\infty]{a.s.} 0.$$
Indeed, for all $i \in \N^{\ast}$, $N_{i} \geq 1$, implying that $\sum\limits_{i=1}^{n} N_{i} \xrightarrow[n\rightarrow\infty]{}\infty$. Moreover by Assumption $\left( {\mbox{H}}_{1}\right)$ and since increments are independent, one gets the following term is finite:
\begin{eqnarray*}
\sum_{n \geq 1} \frac{\mbox{Var}\left(X_{n}\right)}{n^{2}} \; = \; \sum_{n \geq 1}  \sum\limits_{j=1}^{N_{n}}  \left( \Delta t_{nj}\right)^{-2} \mbox{Var} \left[ \Delta D_{nj}\right] \left( \sum\limits_{i=1}^{n} N_{i}\right)^{-2} \; = \; \left[ \frac{\alpha}{\xi^{2}} + \tau^{2}\right] \sum_{n \geq 1}  \sum\limits_{j=1}^{N_{n}}  \left( \Delta t_{nj}\right)^{-1}  \left( \sum\limits_{i=1}^{n} N_{i}\right)^{-2} \; < \; +\infty.
\end{eqnarray*} 
Thus\: $\widehat{m}_{n}^{(1)} \xrightarrow[n\rightarrow\infty]{a.s.} m^{(1)}$. 

\paragraph{{\bf Almost sure convergence of $\widehat{\overline{m}}_{n}^{(2)}$ to $\overline{m}^{(2)}$}}
Let us set:
$$
\tilde{m}_{n}^{(2)} =  \left( \sum\limits_{i=1}^{n} N_{i}\right)^{-1} \sum_{i=1}^{n} \sum_{j=1}^{N_{i}} \left( \Delta t_{ij}\right)^{-1}  \left( \Delta D_{ij} - \E \left[ \Delta D_{ij}\right] \right)^{2}.
$$
Hence the following decomposition holds: $\widehat{\overline{m}}_{n}^{(2)} - \overline{m}^{(2)} = \widehat{\overline{m}}_{n}^{(2)} - \tilde{m}_{n}^{(2)}+ \tilde{m}_{n}^{(2)} - \overline{m}^{(2)}.$ Thus one has to prove that both $\widehat{\overline{m}}_{n}^{(2)} - \tilde{m}_{n}^{(2)}$ and $\tilde{m}_{n}^{(2)} - \overline{m}^{(2)}$ tend almost surely to $0$ as $n$ tends to infinity. 
\begin{enumerate}
\item {\bf Almost sure convergence of $\widehat{\overline{m}}_{n}^{(2)}-\tilde{m}_{n}^{(2)}$ to $0$}
\begin{eqnarray*} 
&&\widehat{\overline{m}}_{n}^{(2)} - \tilde{m}_{n}^{(2)}\\
& = & \left( \sum\limits_{i=1}^{n}N_{i}\right)^{-1} \left[ m^{(1)} - \widehat{m}_{n}^{(1)}\right]  \sum_{i=1}^{n} \sum_{j=1}^{N_{i}} \left[\left(2 \Delta D_{ij} - 2 \E\left( \Delta D_{ij}\right)\right) + \left( \E\left( \Delta D_{ij}\right) - \Delta t_{ij} \widehat{m}_{n}^{(1)}\right)\right]\\
& = & \left[ m^{(1)} - \widehat{m}_{n}^{(1)}\right]^{2} \left( \sum\limits_{i=1}^{n}N_{i}\right)^{-1} \sum_{i=1}^{n} \sum_{j=1}^{N_{i}} \Delta t_{ij} -2 \left[ \widehat{m}_{n}^{(1)} - m^{(1)} \right] \left( \sum\limits_{i=1}^{n}N_{i}\right)^{-1} \sum_{i=1}^{n} \sum_{j=1}^{N_{i}} \left(\Delta D_{ij}-\E\left( \Delta D_{ij}\right)\right).
\end{eqnarray*}
Using Assumption $\left(H_{2}\right)$ and as shown previously one can deduce easily that the first term of the last expression tends to $0$ as $n$ tends to infinity. Moreover the second term tends also to $0$ as $n$ tends to infinity since $\left[ \widehat{m}_{n}^{(1)} - m^{(1)} \right] \xrightarrow[n\rightarrow\infty]{a.s.} 0$ and $\left( \sum\limits_{i=1}^{n}N_{i}\right)^{-1} \sum\limits_{i=1}^{n} \sum\limits_{j=1}^{N_{i}} \left(\Delta D_{ij}-\E\left( \Delta D_{ij}\right)\right) \xrightarrow[n\rightarrow\infty]{a.s.} 0$. Indeed using Lemma \ref{bordes}, Assumption $\left( {\mbox{H}}_{2}\right)$ and since increments are independent, one gets:
\begin{eqnarray*}
 \sum_{n \geq 1}  \sum\limits_{j=1}^{N_{n}} \mbox{Var} \left[ \Delta D_{nj}\right] \left( \sum\limits_{i=1}^{n} N_{i}\right)^{-2} \; = \; \left[ \frac{\alpha}{\xi^{2}} + \tau^{2}\right] \sum_{n \geq 1}  \sum\limits_{j=1}^{N_{n}} \Delta t_{nj}  \left( \sum\limits_{i=1}^{n} N_{i}\right)^{-2} \; < \; \infty.
\end{eqnarray*} 
Thus one can deduce that $\widehat{\overline{m}}_{n}^{(2)} - \tilde{m}^{(2)}_n \xrightarrow[n\rightarrow\infty]{a.s.} 0$. 

\item {\bf Almost sure convergence of $\tilde{m}_{n}^{(2)}$ to $\overline{m}^{(2)}$.}
Applying Theorem \ref{thmpetrov}, it follows that:
$$\tilde{m}_{n}^{(2)} - \overline{m}^{(2)} = \left( \sum\limits_{i=1}^{n} N_{i}\right)^{-1} \sum_{i=1}^{n} \sum_{j=1}^{N_{i}} \left( \left( \Delta t_{ij}\right)^{-1} \left( \Delta D_{ij} - \E \left[ \Delta D_{ij}\right] \right)^{2} - \overline{m}^{(2)}\right)  \xrightarrow[n\rightarrow\infty]{a.s.} 0.$$

Indeed, since increments are independent, one gets that there exists constants $\kappa_{1}\left(\theta \right)$ and $\kappa_{2}\left(\theta \right)$ depend only on $\theta$ (one can compute them explicitly) such that:
\begin{eqnarray*}
\sum_{n \geq 1} \sum_{j=1}^{N_{n}} \left(\Delta t_{nj}\right)^{-2} \mbox{Var} \left[ \left( \Delta D_{nj} - \E \left[ \Delta D_{nj}\right] \right)^{2}\right] \left( \sum\limits_{i=1}^{n} N_{i}\right)^{-2} = \kappa_{1}\left(\theta \right) \sum_{n \geq 1} N_{n}\left( \sum\limits_{i=1}^{n} N_{i}\right)^{-2}  + \kappa_{2}\left(\theta \right) \sum_{n \geq 1} \sum_{j=1}^{N_{n}} \left(\Delta t_{nj}\right)^{-1} \left( \sum\limits_{i=1}^{n} N_{i}\right)^{-2} 
\end{eqnarray*} 
which is finite using Lemma \ref{bordes} and Assumption $\left({\mbox{H}}_{1} \right)$ . Thus it follows that $\tilde{m}_{n}^{(2)}  \xrightarrow[n\rightarrow\infty]{a.s.} \overline{m}^{(2)}$.
\end{enumerate}

\paragraph{{\bf Almost sure convergence of $\widehat{\overline{m}}_{n}^{(3)}$ to $\overline{m}^{(3)}$}}
Similarly as above, we set:
$$\tilde{m}_{n}^{(3)} = \left( \sum\limits_{i=1}^{n} N_{i}\right)^{-1} \sum_{i=1}^{n} \sum_{j=1}^{N_{i}} \left( \Delta t_{ij}\right)^{-1} \left( \Delta D_{ij} - \E \left[ \Delta D_{ij}\right] \right)^{3}.$$

Next we have the following decomposition: $\widehat{\overline{m}}_{n}^{(3)} - \overline{m}^{(3)}=  \widehat{\overline{m}}_{n}^{(3)} - \tilde{m}_{n}^{(3)} + \tilde{m}_{n}^{(3)} - \overline{m}^{(3)}.$ Let us check that $\widehat{\overline{m}}_{n}^{(3)} - \tilde{m}_{n}^{(3)}$ tends almost surely to $0$ as $n$ tends to infinity. 

\begin{enumerate}
\item {\bf Almost sure convergence of $\widehat{\overline{m}}_{n}^{(3)}-\tilde{m}_{n}^{(3)}$ to $0$}.  
\begin{small}
\begin{eqnarray*}
&& \widehat{\overline{m}}_{n}^{(3)} - \tilde{m}_{n}^{(3)} \\
& = & \left[ m^{(1)} - \widehat{m}_{n}^{(1)}\right] \left( \sum\limits_{i=1}^{n}N_{i}\right)^{-1} \sum_{i=1}^{n} \sum_{j=1}^{N_{i}}  \left[ \left[ \Delta D_{ij} - \Delta t_{ij}\widehat{m}_{n}^{(1)}\right]^{2} + \left[ \Delta D_{ij} - \Delta t_{ij}\widehat{m}_{n}^{(1)}\right] \left[\Delta D_{ij} - \E\left(\Delta D_{ij} \right)\right] + \left[ \Delta D_{ij} - \E\left(\Delta D_{ij} \right)\right]^{2}\right]\\
& = &  \left[ m^{(1)} - \widehat{m}_{n}^{(1)}\right] \left( \sum\limits_{i=1}^{n}N_{i}\right)^{-1} \sum_{i=1}^{n} \sum_{j=1}^{N_{i}}  \left[ 3\Delta D_{ij}^{2} + \left( \Delta t_{ij} \hat{m}_{n}^{(1)}\right)^{2} - 3 \Delta t_{ij} \Delta D_{ij} \hat{m}_{n}^{(1)} -3 \Delta D_{ij} \E\left( \Delta D_{ij}\right) + \Delta t_{ij} \hat{m}_{n}^{(1)} \E\left( \Delta D_{ij}\right) + \E\left(\Delta D_{ij}\right)^{2}\right]
\end{eqnarray*}
\end{small}
Let us show that we can replace $\hat{m}_{n}^{(1)}$ by $m^{(1)}$ in the above expression. Using Assumption $\left( H_{2} \right)$, $\E\left( \Delta D_{ij}\right)=\Delta t_{ij} m^{(1)}$ and the fact that $\hat{m}_{n}^{(1)}$ tends to $m^{(1)}$ as $n$ tends almost surely to infinity, it follows that 
\begin{small}
\begin{eqnarray*}
\left[m^{(1)} - \widehat{m}_{n}^{(1)}\right] \left( \sum\limits_{i=1}^{n}N_{i}\right)^{-1} \sum\limits_{i=1}^{n} \sum\limits_{j=1}^{N_{i}} \Delta t_{ij}^{2} \left( \left(\hat{m}_{n}^{(1)}\right)^{2} - \left( m^{(1)}\right)^{2}\right) & = &  -\left[\widehat{m}_{n}^{(1)} - m^{(1)}\right]^{2} \underbrace{\left( \sum\limits_{i=1}^{n}N_{i}\right)^{-1} \sum\limits_{i=1}^{n} \sum\limits_{j=1}^{N_{i}}\Delta t_{ij}^{2} \left[\widehat{m}_{n}^{(1)} + m^{(1)}\right]}_{\leq \: d_{u}^{2}\left(\widehat{m}_{n}^{(1)} + m^{(1)}\right) \xrightarrow[n\rightarrow\infty]{a.s.} 2m^{(1)} d_{u}^{2}} \xrightarrow[n\rightarrow\infty]{a.s.} 0,\\
\left[m^{(1)} - \widehat{m}_{n}^{(1)}\right] \left( \sum\limits_{i=1}^{n}N_{i}\right)^{-1} \sum\limits_{i=1}^{n} \sum\limits_{j=1}^{N_{i}}  \Delta t_{ij} \Delta D_{ij} \left( \hat{m}_{n}^{(1)} - m^{(1)}\right) & = &  -\left[\widehat{m}_{n}^{(1)} - m^{(1)}\right]^{2} \underbrace{\left( \sum\limits_{i=1}^{n}N_{i}\right)^{-1} \sum\limits_{i=1}^{n} \sum\limits_{j=1}^{N_{i}} \Delta t_{ij}^{2} \frac{\Delta D_{ij}}{\Delta t_{ij}}}_{\leq \: d_{u}^{2} \:\hat{m}_{n}^{(1)} \xrightarrow[n\rightarrow\infty]{a.s.} d_{u}^{2}\:m^{(1)}} \xrightarrow[n\rightarrow\infty]{a.s.} 0\\
\left[m^{(1)} - \widehat{m}_{n}^{(1)}\right] \left( \sum\limits_{i=1}^{n}N_{i}\right)^{-1} \sum\limits_{i=1}^{n} \sum\limits_{j=1}^{N_{i}}  \Delta t_{ij}^{2} m^{(1)} \left[\widehat{m}_{n}^{(1)} - m^{(1)}\right] & = & -m^{(1)} \left[\widehat{m}_{n}^{(1)} - m^{(1)}\right]^{(2)} \underbrace{\left( \sum\limits_{i=1}^{n}N_{i}\right)^{-1} \sum\limits_{i=1}^{n} \sum\limits_{j=1}^{N_{i}}  \Delta t_{ij}^{2}}_{\leq\: d_{u}^{2}} \xrightarrow[n\rightarrow\infty]{a.s.} 0.
\end{eqnarray*}
\end{small}

Thus it follows that
\begin{eqnarray*}
\widehat{\overline{m}}_{n}^{(3)} - \tilde{m}_{n}^{(3)}
& = & 3 \left[ m^{(1)} - \widehat{m}_{n}^{(1)}\right] \left( \sum\limits_{i=1}^{n}N_{i}\right)^{-1} \sum_{i=1}^{n} \sum_{j=1}^{N_{i}}\left[\Delta D_{ij}^{2} - \E\left(\Delta D_{ij}\right)^{2} - 2\Delta D_{ij} \E\left(\Delta D_{ij} \right)\right] + o_{a.s.}\left(1 \right) \\
& = &  3 \left[ m^{(1)} - \widehat{m}_{n}^{(1)}\right] \left( \sum\limits_{i=1}^{n}N_{i}\right)^{-1} \sum_{i=1}^{n} \sum_{j=1}^{N_{i}} \left( \Delta D_{ij} - \E\left(\Delta D_{ij} \right) \right)^{2} + o_{a.s.} \left(1 \right) 
\end{eqnarray*}
which tends almost surely to $0$ as $n$ tends to infinity since $$\left( \sum\limits_{i=1}^{n}N_{i}\right)^{-1} \sum\limits_{i=1}^{n} \sum\limits_{j=1}^{N_{i}} \left( \Delta D_{ij} - \E\left(\Delta D_{ij} \right) \right)^{2} \; \leq \; d_{u}\: \tilde{m}_{n}^{(2)} \xrightarrow[n\rightarrow\infty]{a.s.} d_{u}\: \overline{m}^{(2)}$$ and $\widehat{m}_{n}^{(1)} \xrightarrow[n\rightarrow\infty]{as} m^{(1)}$. Then we deduce that $\widehat{\overline{m}}_{n}^{(3)} - \tilde{m}_{n}^{(3)} \xrightarrow[n\rightarrow\infty]{as} 0$.

\item {\bf Almost sure convergence of $\tilde{m}_{n}^{(3)}$ to $\overline{m}^{(3)}$.} 
After tedious calculations, one obtain that there exists constants $\kappa_{3}\left(\theta \right)$, $\kappa_{4}\left(\theta \right)$ and $\kappa_{5}\left(\theta \right)$ depending only on $\theta$ such that:
\begin{eqnarray*}
&& \sum_{n \geq 1} \sum_{j=1}^{N_{n}} \left(\Delta t_{nj}\right)^{-2}  \mbox{Var} \left[ \left( \Delta D_{nj} - \E \left[ \Delta D_{nj}\right] \right)^{3} \right] \left( \sum\limits_{i=1}^{n} N_{i}\right)^{-2}\\
& = & \kappa_{3}\left(\theta \right) \sum\limits_{n \geq 1} \sum\limits_{j=1}^{N_{n}} \left( \Delta t_{nj}\right)^{-1} \left( \sum\limits_{i=1}^{n} N_{i}\right)^{-2} + \kappa_{4}\left(\theta \right) \sum\limits_{n \geq 1} N_{n} \left( \sum\limits_{i=1}^{n} N_{i}\right)^{-2} + \kappa_{5}\left(\theta \right) \sum\limits_{n \geq 1} \sum\limits_{j=1}^{N_{n}} \Delta t_{nj} \left( \sum\limits_{i=1}^{n} N_{i}\right)^{-2}.
\end{eqnarray*} 

All these series, using Lemma \ref{bordes}, $\left( {\mbox{H}}_{1}\right)$ and $\left( {\mbox{H}}_{2}\right)$, are convergent. Thus we have that $\tilde{m}_{n}^{(3)}  \xrightarrow[n\rightarrow\infty]{a.s.} \overline{m}^{(3)}$. 
\hfill$\Box$
\end{enumerate}
\end{pf}

Before showing the asymptotic normality of $\widehat{m}_{n}$, we shall establish the following Lemma.
\begin{lem} \label{lemnorm}
If $(\mbox{H}_{2})$ and the following assumption hold
\begin{itemize}
\item[$(\mbox{H}_{3})$] $\forall u \in \left\lbrace 0, 1, 3 \right\rbrace,\, \displaystyle\lim_{n\rightarrow\infty}\left( \sum\limits_{i=1}^{n} N_{i}\right) ^{-1}\sum\limits_{i=1}^{n} \sum\limits_{j=1}^{N_{i}} \Delta t_{ij}^{u-2} \; =\; c_{u} < \infty$\:.
\end{itemize}
Then it follows that
$$ \left( \sum_{i=1}^{n} N_{i}\right)^{1/2} \begin{pmatrix}
\hat{m}_{n}^{(1)} - m^{(1)} \\
\tilde{m}_{n}^{(2)} - \overline{m}^{(2)}\\
\tilde{m}_{n}^{(3)} - \overline{m}^{(3)}
\end{pmatrix} \xrightarrow[n\rightarrow\infty]{d} N\left(0, \Sigma^{(\infty)}\right),$$
where 
\begin{footnotesize}
$$\Sigma^{(\infty)}\; = \; \begin{pmatrix}
\displaystyle\left(\frac{\alpha}{\xi^{2}} + \tau^{2}\right)c_{1} & \displaystyle\frac{2 \alpha}{\xi^{3}} c_{1} & \displaystyle\frac{6 \alpha}{\xi^{4}} c_{1} +  3\tau^{4} + \frac{6 \alpha \tau^{2}}{\xi^{2}} + \frac{3\alpha^{2}}{\xi^{4}} \\
\displaystyle\frac{2 \alpha}{\xi^{3}} c_{1}  & \displaystyle\frac{6 \alpha}{\xi^{4}}c_{1} +  2\tau^{4} + \frac{4\alpha \tau^{2}}{\xi^{2}} + \frac{2 \alpha^{2}}{\xi^{4}} & \displaystyle\frac{24 \alpha}{\xi^{5}}c_{1} +  \frac{18 \alpha^{2}}{\xi^{5}} + \frac{18 \alpha \tau^{2}}{\xi^{3}}\\
\displaystyle\frac{6 \alpha}{\xi^{4}} c_{1} + 3\tau^{4} + \frac{6 \alpha \tau^{2}}{\xi^{2}} + \frac{3\alpha^{2}}{\xi^{4}}  & \displaystyle \frac{24 \alpha}{\xi^{5}}c_{1} + \frac{18 \alpha^{2}}{\xi^{5}} + \frac{18 \alpha \tau^{2}}{\xi^{3}}& \displaystyle \frac{120 \alpha}{\xi^{6}} c_{1} +  \frac{126 \alpha^{2}}{\xi^{6}} + \frac{90 \alpha \tau^{2}}{\xi^{4}} +\left(\frac{15\alpha^{3}}{\xi^{6}} + \frac{45 \alpha^{2} \tau^{2}}{\xi^{4}} + 15 \tau^{6} + \frac{45\alpha \tau^{4}}{\xi^{2}}\right) c_{3}   
\end{pmatrix}.$$
\end{footnotesize}
\end{lem} 

\begin{pf}
To prove this Lemma we apply the central limit theorem of Lindeberg-Feller \cite{vander} since the increments are independent. We set first by $\Delta \overline{D}_{ij} = \Delta D_{ij} - \E\left( \Delta D_{ij}\right)$. Then we have 
$$ \left( \sum_{i=1}^{n} N_{i}\right)^{1/2} \begin{pmatrix}
\hat{m}_{n}^{(1)} - m^{(1)} \\
\tilde{m}_{n}^{(2)} - \overline{m}^{(2)}\\
\tilde{m}_{n}^{(3)} - \overline{m}^{(3)}
\end{pmatrix} \; = \; \left(\sum_{i=1}^{n} N_{i}\right)^{-1/2} \sum_{i=1}^{n} \sum_{j=1}^{N_{i}} \left( \Delta t_{ij}\right)^{-1} \begin{pmatrix}
\Delta D_{ij} - \E\left( \Delta D_{ij} \right) \\
\Delta \overline{D}_{ij}^{2}  - \E\left( \Delta \overline{D}_{ij}^{2}\right) \\
\Delta \overline{D}_{ij}^{3}  - \E\left( \Delta \overline{D}_{ij}^{3}\right)
\end{pmatrix} \; = \; \left(\sum_{i=1}^{n} N_{i}\right)^{-1/2} \sum_{i=1}^{n} \sum_{j=1}^{N_{i}} \left( \Delta t_{ij}\right)^{-1} X_{ij}.$$

We set $X_{ij} = \left( X_{ij1}, X_{ij2}, X_{ij3}\right)^{T}$. Let us check the first condition of the Lindeberg-Feller theorem. For any $\epsilon > 0$, we have:
\begin{eqnarray}
\nonumber
&&\left( \sum\limits_{i=1}^{n} N_{i} \right)^{-1} \sum_{i=1}^{n} \sum_{j=1}^{N_{i}} \E\left[\frac{\left\| X_{ij}\right\|_{2}^{2}} {\Delta t_{ij}^{2}} \; \displaystyle 1_{\left\lbrace \frac{\left\| X_{ij} \right\|_{2}} {\Delta t_{ij}}  > \epsilon \: \left(\sum\limits_{i=1}^{n} N_{i}\right)^{1/2} \right\rbrace }\right] \; = \; \left( \sum\limits_{i=1}^{n} N_{i} \right)^{-1} \sum_{i=1}^{n} \sum_{j=1}^{N_{i}} \left( \Delta  t_{ij} \right)^{-2}\E\left[\left( \sum\limits_{k=1}^{3} X_{ijk}^{2}\right) \; \displaystyle 1_{\left\lbrace  \frac{ \left\| X_{ij}\right\|_{2}^{2}} {\Delta t_{ij}^{2}}  > \epsilon^{2} \: \sum\limits_{i=1}^{n} N_{i} \right\rbrace }\right]\\
& = & \left( \sum\limits_{i=1}^{n} N_{i} \right)^{-1} \sum\limits_{i=1}^{n} \sum_{j=1}^{N_{i}} \sum\limits_{k=1}^{3} \left( \Delta t_{ij}\right)^{-2} \E\left[\left(  X_{ijk}^{2}\right) \; \displaystyle 1_{\left\lbrace  \frac{ \left\| X_{ij}\right\|_{2}^{2}} {\Delta t_{ij}^{2}}  > \epsilon^{2} \: \left( \sum\limits_{i=1}^{n} N_{i}\right)  \right\rbrace }\right].
\end{eqnarray}
Moreover because
\begin{equation*}
\left\lbrace \sum_{k=1}^{3} X_{ijk}^{2} > \left( \Delta t_{ij} \right)^{2} \epsilon^{2} \left( \sum\limits_{i=1}^{n} N_{i}\right) \right\rbrace \; \subset \; \bigcup\limits_{k=1}^{3} \left\lbrace X_{ijk}^{2} > \frac{\Delta t_{ij}^{2} \epsilon^{2} \left( \sum\limits_{i=1}^{n}N_{i}\right) }{3}\right\rbrace, 
\end{equation*}
we have
\begin{large}
\begin{equation*}
\displaystyle 1_{\left\lbrace \frac{\left\| X_{ij} \right\|_{2}} {\Delta t_{ij}}  > \epsilon \: \left(\sum\limits_{i=1}^{n} N_{i}\right)^{1/2} \right\rbrace } \; \leq \; \displaystyle 1_{ \left\lbrace \bigcup\limits_{k'=1}^{3}\left\lbrace \left|\frac{ X_{ijk'} } {\Delta t_{ij}} \right | > \epsilon \: \left(\sum\limits_{i=1}^{n} N_{i}\right)^{1/2} 3^{-1/2}\right\rbrace \right\rbrace } \; \leq \; \sum\limits_{k'=1}^{3} \displaystyle 1_{ \left\lbrace \left|\frac{ X_{ijk'} } {\Delta t_{ij}} \right | > \epsilon \: \left(\sum\limits_{i=1}^{n} N_{i}\right)^{1/2} 3^{-1/2}\right\rbrace }.
\end{equation*}
\end{large}

Thus it follows that Equation (1) implies that
\begin{eqnarray}
\nonumber
&&\left( \sum\limits_{i=1}^{n} N_{i} \right)^{-1} \sum\limits_{i=1}^{n} \sum_{j=1}^{N_{i}}  \left( \Delta t_{ij}\right)^{-2} \sum\limits_{k=1}^{3} \sum\limits_{k'=1}^{3} \E\left[\left(  X_{ijk}^{2}\right) \; \displaystyle 1_{\left\lbrace \left | \frac{  X_{ijk'}} {\Delta t_{ij}} \right | > \epsilon \: \left(\sum\limits_{i=1}^{n} N_{i} \right)^{1/2} 3^{-1/2}\right\rbrace }\right]\\
\nonumber
& \leq & \sqrt{3} \, \epsilon^{-1} \: \left( \sum\limits_{i=1}^{n} N_{i} \right)^{-1/2} \left( \sum\limits_{i=1}^{n} N_{i} \right)^{-1} \sum\limits_{i=1}^{n} \sum_{j=1}^{N_{i}} \left( \Delta t_{ij}\right)^{-3}\sum\limits_{k=1}^{3} \sum\limits_{k'=1}^{3} \E\left[ X_{ijk}^{2}\left | X_{ijk'}\right| \right]\\
& \leq & \frac{ \epsilon^{-1} \sqrt{3}}{2} \: \left( \sum\limits_{i=1}^{n} N_{i} \right)^{-1/2} \left( \sum\limits_{i=1}^{n} N_{i} \right)^{-1} \sum\limits_{i=1}^{n} \sum_{j=1}^{N_{i}} \left( \Delta t_{ij}\right)^{-3}\sum\limits_{k=1}^{3} \sum\limits_{k'=1}^{3} \left(\E\left[ X_{ijk}^{4} \right] + \E\left[ X_{ijk'}^{2} \right]\right),
\end{eqnarray}
where the last inequality is obtained by applying the Young inequality. Moreover one can check that for any $q \in \N^{*}$, we have
\begin{eqnarray*}
\E\left(\Delta D_{ij}^{q}\right)& = &  \E\left[ \left(\Delta Y_{ij} +\Delta B_{ij}\right)^{q} \right] = \displaystyle{\sum\limits_{s=0}^{q} \binom{q}{s} \:\E \left(\Delta Y_{ij}^{s} \right) \E \left(\Delta B_{ij}^{q-s} \right)} = \displaystyle{\sum\limits_{s=0}^{q} \binom{q}{s} \:\frac{\prod\limits_{l=1}^{s} \left( \alpha \Delta t_{ij} + s - l\right)}{\xi^{s}} \left(\tau^{2}\Delta t_{ij}\right)^{\frac{q-s}{2}} \E \left( \tilde{B}^{q-s}\right)}\\
& = & \displaystyle{\Delta t_{ij}\: \alpha \sum\limits_{s=0}^{q} \binom{q}{s}\: \frac{\prod\limits_{l=1}^{s-1} \left( \alpha \Delta t_{ij} + s - l\right)}{\xi^{s}} \left(\tau^{2} \Delta t_{ij}\right)^{\frac{q-s}{2}} \E \left( \tilde{B}^{q-s}\right)} = \Delta t_{ij} \:\mbox{Pol}_{q-1}\left( \Delta t_{ij}\right),
\end{eqnarray*}
where $\tilde{B}\sim N\left(0,1\right)$ and $\mbox{Pol}_{q-1}\left( \Delta t_{ij}\right)$ denotes a polynomial of order $q-1$ with respect to $\Delta t_{ij}$ the coefficients of which depend only on $\theta$. Then Equation $(2)$ is equal to
\begin{eqnarray*}
&&\frac{\epsilon^{-1}\sqrt{3}}{2} \: \left( \sum\limits_{i=1}^{n} N_{i} \right)^{-1/2} \left( \sum\limits_{i=1}^{n} N_{i} \right)^{-1} \sum\limits_{i=1}^{n} \sum_{j=1}^{N_{i}} \left( \Delta t_{ij}\right)^{-2} \left( \mbox{Pol}_{11}\left( \Delta t_{ij}\right)  + \mbox{Pol}_{5}\left( \Delta t_{ij}\right)\right)\\
& \leq & \frac{\epsilon^{-1} \sqrt{3}}{2} \: \left( \sum\limits_{i=1}^{n} N_{i} \right)^{-1/2} \left( \sum\limits_{i=1}^{n} N_{i} \right)^{-1} \sum\limits_{i=1}^{n} \sum_{j=1}^{N_{i}} \left( \Delta t_{ij}\right)^{-2} \mbox{Pol}_{11}\left( \Delta t_{ij}\right)
\end{eqnarray*}
which tends to 0 as $n$ tends to infinity since $\left( \sum\limits_{i=1}^{n} N_{i} \right)^{-1} \sum\limits_{i=1}^{n} \sum\limits_{j=1}^{N_{i}} \left( \Delta t_{ij}\right)^{-2} \mbox{Pol}_{11}\left( \Delta t_{ij}\right)$ is bounded using Assumptions $\left( H_{2}\right)$ and $\left( H_{3}\right)$.

Next the variance covariance matrix $\Sigma_{ij}$ of $X_{ij}$ is given by
$$
\Sigma_{ij} = \begin{pmatrix}
\mbox{Var}\left( \Delta D_{ij} \right) & \E\left( \Delta \overline{D}_{ij}^{3}\right) & \E\left( \Delta \overline{D}_{ij}^{4}\right)\\
\E\left( \Delta \overline{D}_{ij}^{3}\right) & \mbox{Var}\left[\left(\Delta D_{ij} - \E \left( \Delta D_{ij} \right) \right)^{2}\right] & \E\left( \Delta \overline{D}_{ij}^{5}\right) - \E\left( \Delta \overline{D}_{ij}^{2}\right) \E\left( \Delta \overline{D}_{ij}^{3}\right)\\
\E\left( \Delta \overline{D}_{ij}^{4}\right)& \E\left( \Delta \overline{D}_{ij}^{5}\right) - \E\left( \Delta \overline{D}_{ij}^{2}\right) \E\left( \Delta \overline{D}_{ij}^{3}\right) & \mbox{Var}\left[\left(\Delta D_{ij} - \E \left( \Delta D_{ij} \right) \right)^{3}\right] 
\end{pmatrix}
.$$
Thus the second condition of Lindeberg-Feller theorem is also satisfied since:
$$\mbox{Cov} \left( \left(\sum\limits_{i=1}^{n} N_{i}\right)^{-1/2}\sum\limits_{i=1}^{n} \sum\limits_{j=1}^{N_{i}} \left( \Delta t_{ij}\right)^{-1} \, X_{ij}\right)  \xrightarrow[n\rightarrow\infty]{} \Sigma^{(\infty)},$$
where the finite terms of $\Sigma^{(\infty)}$, under Assumption $\left( {\mbox{H}}_{3}\right) $, are obtained from the following equations:
$$\Sigma^{\left( \infty\right)} \; = \; \lim\limits_{n \rightarrow \infty} \left( \sum\limits_{i=1}^{n} N_{i}\right)^{-1}  \sum\limits_{i=1}^{n} \sum\limits_{j=1}^{N_{i}} \left( \Delta t_{ij}\right)^{-2}\,\Sigma_{ij}$$ 
such that $\sigma_{uv}$, pour $1 \leq u \leq v \leq 3$, are the terms of the variance-covariance matrix $\Sigma_{ij}$:
\begin{eqnarray*}
\sigma_{11} & = & \left(\frac{\alpha}{\xi^{2}} + \tau^{2}\right)\Delta t_{ij}, \;\sigma_{12}\;=\;  \frac{2 \alpha}{\xi^{3}}\Delta t_{ij}, \; \sigma_{13} \; = \; \frac{6 \alpha}{\xi^{4}} \Delta t_{ij} + \left( 3\tau^{4} + \frac{6 \alpha \tau^{2}}{\xi^{2}} + \frac{3\alpha^{2}}{\xi^{4}}\right)\Delta t_{ij}^{2} \\
\sigma_{22} & = & \frac{6 \alpha}{\xi^{4}}\Delta t_{ij} + \left( 2\tau^{4} + \frac{4\alpha \tau^{2}}{\xi^{2}} + \frac{2 \alpha^{2}}{\xi^{4}}\right)\Delta t_{ij}^{2}, \; \sigma_{23} \; = \; \frac{24 \alpha}{\xi^{5}}\Delta t_{ij} + \left( \frac{18 \alpha^{2}}{\xi^{5}} + \frac{18 \alpha \tau^{2}}{\xi^{3}} \right)\Delta t_{ij}^{2}\\
\sigma_{33} & = & \frac{120 \alpha}{\xi^{6}}\Delta t_{ij} + \left( \frac{126 \alpha^{2}}{\xi^{6}} + \frac{90 \alpha \tau^{2}}{\xi^{4}}\right) \Delta t_{ij}^{2}+\left(\frac{15\alpha^{3}}{\xi^{6}} + \frac{45 \alpha^{2} \tau^{2}}{\xi^{4}} + 15 \tau^{6} + \frac{45\alpha \tau^{4}}{\xi^{2}}\right)\Delta t_{ij}^{3}.
\end{eqnarray*}

Finally we conclude that
$$\left(\sum_{i=1}^{n} N_{i}\right)^{-1/2} \sum_{i=1}^{n} \sum_{j=1}^{N_{i}} \left( \Delta t_{ij}\right)^{-1} X_{ij} \xrightarrow[n\rightarrow\infty]{d} N\left(0, \Sigma^{(\infty)}\right).$$
\end{pf}

In the sequel we will prove the asymptotic normality of $\widehat{\theta}_{n}$. First, let us prove the asymptotic normality of $\widehat{m}_{n}$.  
\begin{thm}\label{theodelta}
Under Assumptions $\left({\mbox{H}}_{1}-{\mbox{H}}_{3}\right)$, we have:
$$\left( \sum_{i=1}^{n} N_{i}\right)^{1/2} \left(\widehat{m}_{n} - m \right)  \xrightarrow[n\rightarrow\infty]{d} N\left( 0, H \right),$$
where $H = \mbox{A} \,\,\Sigma^{(\infty)}\,\, \mbox{A}^{T}$ such that ${\mbox{A}}$ is given by:

$${\mbox{A}} = \begin{pmatrix}
\displaystyle 1 & \displaystyle 0 & \displaystyle 0 \\
\displaystyle 0 & \displaystyle 1 & \displaystyle 0 \\
\displaystyle - 3 \left(\frac{\alpha}{\xi^{2}} + \tau^{2}\right)c_{3}  & \displaystyle 0 & \displaystyle 1 
\end{pmatrix}.$$
\end{thm}

\begin{pf}
First we note that
\begin{eqnarray*}
\left( \sum_{i=1}^{n} N_{i}\right)^{1/2} \left(\hat{m}_{n} - m \right)& = & \left( \sum_{i=1}^{n} N_{i}\right)^{1/2} \begin{pmatrix}
\hat{m}_{n}^{(1)} - m^{(1)} \\
\hat{\overline{m}}_{n}^{(2)} - \tilde{m}_{n}^{(2)} +  \tilde{m}_{n}^{(2)} - \overline{m}^{(2)}\\
\hat{\overline{m}}_{n}^{(3)} - \tilde{m}_{n}^{(3)} +  \tilde{m}_{n}^{(3)} - \overline{m}^{(3)}
\end{pmatrix}\; = \; \left( \sum_{i=1}^{n} N_{i}\right)^{1/2}\begin{pmatrix}
0\\
\hat{\overline{m}}_{n}^{(2)} - \tilde{m}_{n}^{(2)}\\
\hat{\overline{m}}_{n}^{(3)} - \tilde{m}_{n}^{(3)}
\end{pmatrix} \; + \; \left( \sum_{i=1}^{n} N_{i}\right)^{1/2} \begin{pmatrix}
\hat{m}_{n}^{(1)} - m^{(1)} \\
\tilde{m}_{n}^{(2)} - \overline{m}^{(2)}\\
\tilde{m}_{n}^{(3)} - \overline{m}^{(3)}
\end{pmatrix}.
\end{eqnarray*} 

Second we have
\begin{eqnarray}
&&\left( \sum_{i=1}^{n} N_{i}\right)^{1/2} \left( \hat{\overline{m}}_{n}^{(2)} - \tilde{m}_{n}^{(2)}\right)\\
\nonumber
& = & 2 \left( \sum_{i=1}^{n} N_{i}\right)^{1/2} \left[m^{(1)} - \hat{m}_{n}^{(1)}\right] \left( \sum_{i=1}^{n} N_{i}\right)^{-1} \sum_{i=1}^{n} \sum_{j=1}^{N_{i}} \left(\Delta D_{ij} - \E\left( \Delta D_{ij}\right) \right) + \left(\sum_{i=1}^{n}N_{i}\right)^{-1} \sum_{i=1}^{n} \sum_{j=1}^{N_{i}} \Delta t_{ij}  \left( \sum_{i=1}^{n} N_{i}\right)^{1/2} \left[ m^{(1)} - \hat{m}_{n}^{(1)} \right]^{2} 
\end{eqnarray}
which tends to 0 as $n$ tends to infinity. Indeed we check that the first term of the last expression tends in probability to $0$ as $n$ tends to infinity since $\left( \sum\limits_{i=1}^{n} N_{i}\right)^{1/2} \left[m^{(1)} - \hat{m}_{n}^{(1)}\right]$ is normally distributed and, as shown previously, $\left( \sum\limits_{i=1}^{n} N_{i}\right)^{-1} \sum\limits_{i=1}^{n} \sum\limits_{j=1}^{N_{i}} \left(\Delta D_{ij} - \E\left( \Delta D_{ij}\right) \right)$ tends to 0 as $n$ tends to infinity. Moreover the second term in the right-hand side of (3) tends to $0$ as $n$ tends to infinity because $\left(\sum\limits_{i=1}^{n}N_{i}\right)^{-1} \sum\limits_{i=1}^{n} \sum\limits_{j=1}^{N_{i}} \Delta t_{ij}$ is convergent, $\left( \sum\limits_{i=1}^{n} N_{i}\right)^{1/2} \left[\hat{m}_{n}^{(1)} - m^{(1)}\right]$ has an asymptotic normal distribution and $\left[\hat{m}_{n}^{(1)} - m^{(1)}\right]$ tends almost surely to 0 as $n$ tends to infinity. Then we deduce that $\left( \sum\limits_{i=1}^{n} N_{i}\right)^{1/2} \left( \hat{\overline{m}}_{n}^{(2)} - \tilde{m}_{n}^{(2)}\right)$ tends in probability to 0 as $n$ tends to infinity.
 
Furthermore one gets that
\begin{eqnarray*}
\left( \sum\limits_{i=1}^{n}N_{i}\right)^{1/2} \left( \hat{\overline{m}}_{n}^{(3)} - \tilde{m}_{n}^{(3)}\right)& = & 3 \left( \sum\limits_{i=1}^{n}N_{i}\right)^{1/2}\left[ m^{(1)} - \widehat{m}_{n}^{(1)}\right] \left( \sum\limits_{i=1}^{n}N_{i}\right)^{-1} \sum_{i=1}^{n} \sum_{j=1}^{N_{i}} \left[ \left( \Delta D_{ij} - \E\left(\Delta D_{ij} \right) \right)^{2} \right] + o_{p} \left(1 \right). 
\end{eqnarray*}
Let us show that $$\left( \sum\limits_{i=1}^{n}N_{i}\right)^{-1} \sum\limits_{i=1}^{n} \sum\limits_{j=1}^{N_{i}} \left[ \left( \Delta D_{ij} - \E\left(\Delta D_{ij} \right) \right)^{2} - \mbox{Var}\left( \Delta D_{ij}\right)\right] \xrightarrow[n\rightarrow\infty]{a.s.}0.$$ Indeed, since increments are independent, one gets that there exists constants $\kappa_{1}\left(\theta \right)$ and $\kappa_{2}\left(\theta \right)$ depending only on $\theta$ such that
\begin{eqnarray*}
\sum_{n \geq 1} \sum_{j=1}^{N_{n}} \mbox{Var} \left[ \left( \Delta D_{nj} - \E \left[ \Delta D_{nj}\right] \right)^{2}\right] \left( \sum\limits_{i=1}^{n} N_{i}\right)^{-2} & = & \kappa_{1}\left(\theta \right) \sum_{n \geq 1} \sum_{j=1}^{N_{n}}\Delta t_{nj}\left( \sum\limits_{i=1}^{n} N_{i}\right)^{-2}  + \kappa_{2}\left(\theta \right) \sum_{n \geq 1} \sum_{j=1}^{N_{n}} \left(\Delta t_{nj}\right)^{2} \left( \sum\limits_{i=1}^{n} N_{i}\right)^{-2}
\end{eqnarray*} 
which is convergent using Assumption $\left( H_{2} \right)$ and Lemma \ref{bordes}. Moreover we have: $$\left( \sum\limits_{i=1}^{n}N_{i}\right)^{-1} \sum\limits_{i=1}^{n} \sum\limits_{j=1}^{N_{i}} \mbox{Var} \left( \Delta D_{ij} \right) \xrightarrow[n\rightarrow\infty]{} \left(\frac{\alpha}{\xi^{2}} + \tau^{2}\right)c_{3}.$$
Then one can write that 

\begin{eqnarray*}
\left( \sum_{i=1}^{n} N_{i}\right)^{1/2} \begin{pmatrix}
\hat{m}_{n}^{(1)} - m^{(1)} \\
\hat{\overline{m}}_{n}^{(2)} - \overline{m}^{(2)}\\
\hat{\overline{m}}_{n}^{(3)} - \overline{m}^{(3)}
\end{pmatrix} & = &
\left(\sum_{i=1}^{n} N_{i}\right)^{-1/2} \sum_{i=1}^{n} \sum_{j=1}^{N_{i}} \left( \Delta t_{ij}\right)^{-1}\begin{pmatrix}
1 & 0 & 0 \\
0 & 1 & 0 \\
- 3 \left(\frac{\alpha}{\xi^{2}} + \tau^{2}\right)c_{3}  & 0 & 1 
\end{pmatrix} \begin{pmatrix}
\Delta D_{ij} - \E\left( \Delta D_{ij} \right) \\
\Delta \overline{D}_{ij}^{2}  - \E \left( \Delta \overline{D}_{ij}\right)^{2} \\
 \Delta \overline{D}_{ij}^{3}  - \E \left( \Delta \overline{D}_{ij}\right)^{3} 
\end{pmatrix} + o_{p}\left(1 \right)\\ 
& = & A \: \left(\sum\limits_{i=1}^{n} N_{i}\right)^{-1/2}\sum_{i=1}^{n} \sum_{j=1}^{N_{i}} \left( \Delta t_{ij}\right)^{-1} \, X_{ij}\;.
\end{eqnarray*}

By Lemma \ref{lemnorm} we have $\left(\sum\limits_{i=1}^{n} N_{i}\right)^{-1/2} \sum\limits_{i=1}^{n} \sum\limits_{j=1}^{N_{i}} \left( \Delta t_{ij}\right)^{-1} X_{ij} \xrightarrow[n\rightarrow\infty]{d} N\left(0, \Sigma^{(\infty)}\right)$. Thus it follows that
$$\left( \sum\limits_{i=1}^{n} N_{i}\right)^{1/2} \left(\widehat{m}_{n} - m \right)  \xrightarrow[n\rightarrow\infty]{d} N\left( 0, \mbox{A} \,\,\Sigma^{(\infty)}\,\, \mbox{A}^{T} \right).$$
\end{pf}

Since $f$ is a differentiable and bijective function and $f^{-1}$ is continuous on $f(\Theta)$, then we obtain the asymptotic normality of $\widehat{\theta}_{n}$ by applying the $\delta$-method (see Theorem $3.1$ in \cite{vander}).
\begin{thm}\label{theodelta1}
Under Assumptions $\left( {\mbox{H}}_{1}-{\mbox{H}}_{3}\right)$, we have:
$$\left( \sum\limits_{i=1}^{n} N_{i}\right)^{1/2}\left( \widehat{\theta}_{n} - \theta\right)  \xrightarrow[n\rightarrow\infty]{d} N\left( 0, M\right),$$
where $M =  G\, H\, G^{T}$ such that $G$ is given by:
$$ G = \begin{pmatrix}
\displaystyle\frac{\partial f_{1}^{-1}}{\partial m^{(1)}} & \displaystyle\frac{\partial f_{2}^{-1}}{\partial m^{(1)}} & \displaystyle\frac{\partial f_{3}^{-1}}{\partial m^{(1)}}\\
\displaystyle\frac{\partial f_{1}^{-1}}{\partial \overline{m}^{(2)}} & \displaystyle\frac{\partial f_{2}^{-1}}{\partial \overline{m}^{(2)}} & \displaystyle\frac{\partial f_{3}^{-1}}{\partial \overline{m}^{(2)}}\\
\displaystyle\frac{\partial f_{1}^{-1}}{\partial{\overline{m}^{(3)}}} & \displaystyle\frac{\partial f_{2}^{-1}}{\partial \overline{m}^{(3)}} & \displaystyle\frac{\partial f_{3}^{-1}}{\partial \overline{m}^{(3)}}
\end{pmatrix} \left(m \right)\; = \; \begin{pmatrix}
\displaystyle \frac{1}{\sqrt{2 m^{(1)} \overline{m}^{(3)}}} & \displaystyle \sqrt{\frac{2m^{(1)}}{\overline{m}^{(3)}}} + \sqrt{\frac{m^{(1)}}{2\overline{m}^{(3)}}} & \displaystyle -\frac{1}{2}\sqrt{\frac{\overline{m}^{(3)}}{2m^{(1)}}}\\
\displaystyle 0 & \displaystyle 0 & \displaystyle 1\\
\displaystyle - \frac{1}{\overline{m}^{(3)}}\sqrt{\frac{m^{(1)}}{2\overline{m}^{(3)}}} & \displaystyle - \frac{m^{(1)}}{\overline{m}^{(3)}} \sqrt{\frac{m^{(1)}}{2\overline{m}^{(3)}}} & \displaystyle -\frac{1}{2}\sqrt{\frac{m^{(1)}}{2\overline{m}^{(3)}}}
\end{pmatrix}.$$
\hfill$\Box$
\end{thm}

\subsection{Statistical inference}

As an application of theorem \ref{theodelta1}, one can construct the confidence interval with asymptotic level $1-\vartheta$ for each parameter:
$$
\lim\limits_{n \rightarrow \infty} \mbox{P}\left(\xi \in \left[ \widehat{\xi}_{n} \pm z_{1-\frac{\vartheta}{2}}\frac{\sigma\left( \widehat{\xi}_{n}\right) }{\sqrt{\sum_{i=1}^{n}N_{i}} } \right] \right) = 1- \vartheta,\;\; \lim\limits_{n \rightarrow \infty} \mbox{P}\left(\alpha \in \left[ \widehat{\alpha_{n}} \pm z_{1-\frac{\vartheta}{2}}\frac{\sigma\left( \widehat{\alpha_{n}}\right) }{\sqrt{\sum_{i=1}^{n}N_{i}}} \right] \right) = 1- \vartheta$$
and
$$\lim\limits_{n \rightarrow \infty} \mbox{P}\left(\tau^{2} \in \left[ \widehat{\tau}^{2}_{n} \pm z_{1-\frac{\vartheta}{2}}\frac{\sigma\left( \widehat{\tau}^{2}_{n}\right) }{\sqrt{\sum_{i=1}^{n}N_{i}}} \right] \right) = 1- \vartheta\\
$$
where $z_{1-\frac{\vartheta}{2}}$ is the critical value of the standard normal distribution and $\sigma\left( \widehat{\xi}_{n}\right), \sigma\left( \widehat{\alpha}_{n}\right)$ and $\sigma\left( \widehat{\tau}_{n}^{2}\right)$ are the asymptotic standard deviation of $\widehat{\xi}_{n}, \widehat{\alpha}_{n}$ and $\widehat{\tau}_{n}^{2}$ respectively (square-root of the diagonal of variance-covariance matrix $M$ appeared in Theorem \ref{theodelta1}). Thus one can test whether $\tau^{2} = 0$ or not, that could be important to determine if the model is a gamma process or not. Moreover, applying the $\delta$-method and using the previous theorem, it can be proved that $$\sqrt{n}\left( \frac{\widehat{\alpha}_{n}}{\widehat{\xi}_{n}^{2}} - \frac{\alpha}{\xi^{2}}\right) \xrightarrow[n\rightarrow\infty]{d} N\left(0,G_{1}\,{\mbox{M}}_{1}\,G_{1}^{T}\right),$$ where ${\mbox{M}}_{1}$ is the top left $2\times 2$ block matrix of $M$ and where $G_{1} = \left(-\frac{2 \alpha}{\xi^{3}}, \frac{1}{\xi^{2}} \right)^{T} $. Hence one can obtain the confidence interval with asymptotic level $1-\vartheta$ for $\alpha/\xi^{2}$. As mentioned in the introduction, it is useful to test the Brownian motion with a positive drift model against the gamma process model.

\section{Particular cases}

Before considering several particular cases corresponding to various sampling scheme, we will introduce some stronger but more comprehensive assumptions: 
\begin{itemize}
\item[$(\mbox{A}_{1})$] Same number of observations for all the processes: $\forall i \in \N^{\ast}, N_i = N$
\item[$(\mbox{A}_{2})$] Same instants of observations for all the processes: $\forall i \in \N^{\ast}, \forall j \in \left\lbrace 1, \ldots, N_{i}\right\rbrace, t_{ij} = t_j$
\item[$(\mbox{A}_{3})$] Regular instants (not necessary the same instants for all the processes): 
$
\forall i \in \N^{\ast}, \exists T_i$ such that $\forall j \in \left\lbrace 1, \ldots, N_{i}\right\rbrace, \Delta t_{ij} = T_i/N_i
$
\item[$(\mbox{A}_{4})$] Same time interval for observations: $\exists T$ such that $\forall i \in \N^{\ast}, t_{iN_i} \leq T$
\item[$(\mbox{A}_{5})$] Uniformly bounded delay between consecutive observations:
$
\exists d_{l}>0,  \forall i \in \N^{\ast}, \forall j \in \left\lbrace 1, \ldots, N_{i}\right\rbrace , \Delta t_{ij} \geq d_{l}
$
\end{itemize}
Note that $({\mbox{A}}_2) \Rightarrow ({\mbox{A}}_1)$. More interesting are the relationships between Assumptions $({\mbox{H}}_1-{\mbox{H}}_3)$ and Assumptions $({\mbox{A}}_1-{\mbox{A}}_5)$. In particular, one can easily check that $({\mbox{A}}_4) \Rightarrow ({\mbox{H}}_2)$ and $({\mbox{A}}_1)\, \& \,({\mbox{A}}_5) \Rightarrow ({\mbox{H}}_1)$.
Moreover simplifications may occur under some assumptions. For instance, if $({\mbox{A}}_3)$ and $({\mbox{A}}_4)$ are satisfied, then $({\mbox{H}}_1)$ and $({\mbox{H}}_3)$ are equivalent respectively to: 
\begin{itemize}
\item[$(\mbox{H}_{1}')$] \; $\sum\limits_{n \geq 1}N_{n}^{2} \left( \sum\limits_{i=1}^{n} N_{i}\right)^{-2}  < \infty$
\item[$(\mbox{H}_{3}')$] \; $\forall u \in \left\lbrace 0, 1\right\rbrace, \: \displaystyle\lim_{n\rightarrow\infty} \left( \sum\limits_{i=1}^{n}N_{i}\right)^{-1} \left( \sum\limits_{i=1}^{n}N_{i}^{3-u}\right)< \infty $
\end{itemize}
In addition if $(\mbox{A}_{1})$ holds then $(\mbox{H}_{1}')$ and $(\mbox{H}_{3}')$ are satisfied. We will now consider five different special cases that can be described in terms of Assumptions $({\mbox{A}}_1-{\mbox{A}}_5)$:
\begin{itemize}
\item Case 1 - Same number of observations a the same regular instants over $[0,T]$: $({\mbox{A}}_1) - ({\mbox{A}}_4)$;
\item Case 2 - Same number of observations at the same non-regular instants over $[0,T]$ : $({\mbox{A}}_1)$, $({\mbox{A}}_2)$ and $({\mbox{A}}_4)$;
\item Case 3 - $N_i=i$ and regular instants over $[0,T]$: $({\mbox{A}}_3)$ and $({\mbox{A}}_4)$;
\item Case 4 - $N_i=i$ and regular instants over $[0,iT]$: $({\mbox{A}}_3)$ and $({\mbox{A}}_5)$;
\item Case 5 - $N_i=2^{i-1}$ and regular instants over $[0,T]$: $({\mbox{A}}_3)$ and $({\mbox{A}}_4)$.
\end{itemize}
One can easily check that estimators in cases 1, 2 and 4 are consistent and asymptotically normal. At least, the estimator in case 3 is consistent but asymptotic normality cannot be established using our results. In the last case one can check that consistency and asymptotic normality cannot be established using our results.

\section{Numerical illustration}
Here we illustrate our theoretical results throughout simulations. We recall that the parameters were fixed as follows: $\xi=1$, $\alpha =0.02$ and $\tau^2=0.02$. The number of observations for each item was set to $N=3$ instants between $0$ and $T=1000$ such that $\Delta t_{i1} = 200, \Delta t_{i2} = 300$ and $\Delta t_{i3} = 500$. We have computed the empirical bias, the empirical squared error (MSE) and the empirical standard deviation (StD) for 1000 repetitions. Table \ref{tab:avbias}, Table \ref{tab:avmse} and Table \ref{tab:avStD} report respectively the empirical bias and the empirical standard deviation for several sample sizes $n$. Based on results given in the Tables \ref{tab:avbias}, \ref{tab:avmse} and \ref{tab:avStD} we note the the average of degradation is well estimated whatever the sample size since the larger $n$ is, the better the estimation is towards Bias, MSE and StD.
\begin{table}[h!]
\begin{minipage}[t]{.46\linewidth}
\caption{Empirical bias}
\label{tab:avbias}
\begin{center}
    \begin{tabular}{c cccc}
\hline
Bias & 50 & 100 & 200 \\
\hline
$\xi$ &  2.22e-1 & 1.44e-1 & 6.25e-2  \\ 
$\alpha$ &  5.55e-3 & 3.61e-3 & 1.57e-3 \\
$\tau^{2}$ &  6.21e-3 & 1.16e-3 & 5.01e-4 \\ 
\hline
1000 repetitions & 937 & 983 & 998\\
\hline
\end{tabular} 
\end{center}
\end{minipage}
\hfill
\begin{minipage}[t]{.46\linewidth}
\caption{Empirical MSE}
\label{tab:avmse}
\begin{center}
   \begin{tabular}{c cccc}
\hline
MSE & 50 & 100 & 200 \\
\hline
$\xi$ & 8.29e-1 & 3.35e-1 & 7.07e-2 \\
$\alpha$ &  5.25e-4 & 2.12e-4 & 4.56e-5 \\
$\tau^2$ & 9.67e-5 & 5.88e-5 & 3.01e-5 \\
\hline
1000 repetitions & 937 & 983 & 998\\
\hline
\end{tabular} 
\end{center}
\end{minipage}
\end{table}

\begin{table}
\caption{Empirical standard deviation}
\label{tab:avStD}
\begin{center}
   \begin{tabular}{c cccc}
\hline
StD & 50 & 100 & 200 \\
\hline
$\xi$ & 6.32e-1 & 5.61e-1 & 5.21e-1 \\
$\alpha$ &  7.78e-1 & 7.48e-1 & 7.41e-1 \\
$\tau^2$ & 0.11e1 & 5.01e-1 & 2.12e-1 \\
\hline
1000 repetitions & 937 & 983 & 998\\
\hline
\end{tabular} 
\end{center}
\end{table}

\section{Real data application}

In what follows, we present the results that we have achieved in the implementation of the  data given in the following sections:

\subsection{NIST dataset}

An example of dataset can be found in \cite{Hand}. Fifteen components were tested under three different temperatures $65^{\circ}C$, $85^{\circ}C$ and $105^{\circ}C$. Degradation percent values were read out at $200, 500$ and $1000$ hours. We have estimated the three parameters of the degradation models and we have constructed, see Table \ref{tab:Résultats}, the $95 \%$ confidence interval of each parameter. First, we denote that values within brackets constitute the standard deviation of each parameter.
\begin{table}[htbp]
\caption{Estimation of parameters and $95 \%$ confidence intervals}
\label{tab:Résultats}
\begin{center}
\begin{tabular}{ l  c  c  c  c }\hline
Parameters & $\xi$ & $\alpha$ & $\tau^{2}$ & $\alpha/\xi^{2}$\\ \hline
Estimation \:$\left(65^{\circ} C \right) $ & $5.51\:(1.31)$ & $0.01\:(0.002)$ & $0.0001\:(19.96)$ & $0.0006$\\ 
Confidence intervals $\left( 95 \%\right) $& $\left[4.84; 6.18\right]$  & $\left[0.01;0.02\right]$ & $\left[0;10.11\right]$ & $\left[0.0004;0.0007\right]$ \\ 
Estimation \:$\left(85^{\circ} C \right) $ & $0.71 \:(0.37)$ & $0.012 \:(0.49)$ & $0.0068 \:(0.111)$ & $0.025$ \\ 
Confidence intervals $\left( 95 \%\right) $& $\left[0.51;0.89\right]$  & $\left[ 0;0.26\right]$ & $\left[ 0;0.06\right]$ & $\left[0;0.54\right]$ \\
Estimation \:$\left(105^{\circ} C \right) $ & $0.29 \:(1.87)$ & $0.02 \:(0.14)$ & $0.27\:(1.51)$ &  $0.25$\\ 
Confidence intervals $\left( 95 \%\right) $& $\left[0; 1.24\right]$  & $\left[ 0;0.09\right]$ & $\left[0;1.04\right]$ & $\left[0;1.06\right]$ \\ \hline
\end{tabular}
\end{center}
\end{table}
Let us discuss the results. One can note that $\xi$ decreases as temperature increases. Moreover $\tau^{2}$ and $\alpha/\xi^2$ increase as temperature increases. However $\alpha$ is almost stable. Finally from the confidence intervals at $65^{\circ} C$ our model turns to be a gamma process since one can accept that $\tau^2 = 0$ and $\alpha/\xi^2 \neq 0$.

\subsection{Heating cable test data}

Whitmore and Schenkelberg \cite{Whi} presented some heating cable test data. The degradation of the cable is measured as the natural logarithm of resistance. Degradation is accelerated by thermal stress so temperature is used as the stress measure. Five test items were baked in an oven at each test temperature. Three test temperatures were used, $200^ {\circ}C$, $240^{\circ}C$ and $260^{\circ}C$, giving a total of $15$ items. The clock times are in thousands of hours. The cable is deemed to have failed when the log-resistance reaches $\ln(2) = 0.693$. The test continued at the lowest test temperature $200^ {\circ}C$ until the test equipment was required for other projects. We have estimated the three parameters of the degradation models and we have constructed, see Table \ref{tab:Résultats1}, the $95 \%$ confidence interval of each parameter. Like above, we denote that values within brackets constitute the standard deviation of each parameter.
\begin{table}[htbp]
\caption{Estimation of parameters and $95 \%$ confidence intervals}
\label{tab:Résultats1}
\begin{center}
\begin{tabular}{ l  c  c  c  c }\hline
Parameters & $\xi$ & $\alpha$ & $\tau^{2}$ & $\alpha/\xi^{2}$\\ \hline 
Estimation \:$\left(200^{\circ} C \right)$ & $2.18 \:(5.07)$ & $0.47 \:(1.41)$ & $0.03 \:(9.97)$ &  $0.09$\\ 
Confidence intervals $\left( 95 \%\right) $& $\left[0.77;3.58\right]$  & $\left[0.08;0.86\right]$ & $\left[0;2.81\right]$ & $\left[0;0.21\right]$ \\ 
Estimation \:$\left(240^{\circ} C \right) $ & $2.38 \:(6.74)$ & $2.17 \:(3.10)$ & $0.14 \:(11.57)$ & $0.38$ \\ 
Confidence intervals $\left( 95 \%\right) $& $\left[0.51;4.25\right]$  & $\left[ 1.31;3.03\right]$ & $\left[ 0;3.35\right]$ & $\left[0;0.88\right]$ \\ 
Estimation \:$\left(260^{\circ} C \right) $ & $2.75 \:(9.01)$ & $5.13 \:(3.78)$ & $0.04 \:(17.31)$ &  $0.67$\\  
Confidence intervals $\left( 95 \%\right) $& $\left[0;5.74\right]$  & $\left[3.87;6.38\right]$ & $\left[0;5.78\right]$ & $\left[0;2.01\right]$ \\ \hline
\end{tabular}
\end{center}
\end{table}
One notes that $\xi$, $\alpha$ and $\alpha/\xi^2$ increase as temperature increases. However it is not the case for $\tau^{2}$. Although we have the same number of items as for the previous data set, here we observe standard deviations with very large values. It is therefore difficult to choose between one of the two sub-models, and more generally it may be interpreted as bad fitting of the model.

\section{Concluding remarks}
In this paper we have proposed a gamma process perturbed by a Brownian motion as a degradation model for which we derived parameters estimator. Asymptotic properties of this estimator have been established. Since degradation of system is also influenced by the environment, it is interesting to consider a model integrating covariates. Such model will be studied in a forthcoming paper.

\section*{References}
\bibliographystyle{elsarticle-num}

\end{document}